\documentstyle[preprint,aps]{revtex}

\begin{document}

\draft
%\preprint{}  
\title{Spontaneous Breaking of Flavor Symmetry and Naturalness \\ 
of Nearly Degenerate Neutrino Masses and Bi-maximal Mixing\footnote{
To appear in SCIENCE IN CHINA (Series A), Vol.35  No.9 (2000)} }
\author{ Yue-Liang  Wu\footnote{Supported by Outstanding Young Scientist Research Fund.} }
\address{Institute of Theoretical Physics, Chinese Academy of Sciences, Beijing 100080, China } 
\date{ylwu@itp.ac.cn}
\maketitle

\begin{abstract} 
The gauge model with $SO(3)_{F}$ flavor symmetry and three Higgs triplets is studied.
We show how the intriguing nearly degenerate neutrino mass and bi-maximal mixing 
scenario comes out naturally after spontaneous breaking of the symmetry. The hierarchy 
between the neutrino mass-squared differences, which is needed for reconciling both solar and 
atmospheric neutrino data, is naturally resulted from an approximate permutation symmetry.
The model can also lead to interesting phenomena on lepton-flavor violations via the 
$SO(3)_{F}$ gauge interactions.  
\vspace{1.5cm}
\end{abstract}


%\narrowtext

 The standard model (SM) has been tested by more and more precise experiments, 
its greatest success is the gauge symmetry structure $SU(3)_{c}\times SU_{L}(2)\times U_{Y}(1)$. 
While neutrinos are assumed to be massless in the SM. Super-Kamiokande 
collaboration\cite{SUPERK1,SUPERK2,SUPERK3,SUPERK4} 
has reported the evidences for oscillation of atmospheric neutrinos 
and for the deficit of the measured solar neutrino flux that is only about 
half of that expected from the `BP' standard 
solar model\cite{BP}. This strongly suggests non-zero neutrino masses.   
Massive neutrinos are also regarded as the best candidate for hot dark matter and may 
play an essential role in the evolution of the large-scale structure 
of the universe\cite{HDM}. To introduce neutrino masses and mixings, it is necessary to modify 
and go beyond the SM. In the recent papers\cite{YLW1,YLW2,YLW0}, we have introduced the gauged 
$SO(3)_{F}$ flavor symmetry\cite{SO3,SO31,SO32,SO33,SO34,SO35}\footnote{See also, 
R. Barbieri, L.J. Hall, G.L. Kane and G.G. Ross, hep-ph/9901228; 
A. Ghosal, hep-ph/9905470.} to describe the three lepton families. 
Some remarkable features have been found to be applicable to the current interesting phenomena of 
neutrinos. After a detailed analysis on various possible scenarios, we have shown that 
the nearly degenerate neutrino mass and bi-maximal mixing scenario\cite{BMM} is the most 
favorable one in our model\cite{YLW2} to reconcile both solar and atmospheric neutrino flux 
anomalies. But its origin has not explicitly been explored in the refs.\cite{YLW1,YLW2,YLW0}. 
To understand the naturalness of the scenario, we will study in this paper the $SO(3)_{F}$ 
gauge model with three Higgs triplets and pay attention to the spontaneous breaking of the 
$SO(3)_{F}$ flavor symmetry in the Higgs sector. As a consequence, the  
nearly degenerate neutrino mass and bi-maximal mixing scenario is naturally derived after 
spontaneous symmetry breaking. 

 For a less model-dependent analysis, we directly start from an $SO(3)_{F}\times 
SU(2)_{L}\times U(1)_{Y}$ invariant effective lagrangian with three $SO(3)_{F}$ Higgs triplets 
\begin{eqnarray}
& & {\cal L} =   {\cal L}_{SM}  + \frac{1}{2}g'_{3}A_{\mu}^{k}
( \bar{L}_{i}\gamma^{\mu} (t^{k})_{ij}L_{j} 
+ \bar{e}_{R i} \gamma^{\mu}(t^{k})_{ij}e_{R j} ) \nonumber \\
& & + (c_{1}\varphi_{i}\varphi_{j}\chi +  c'_{1} \varphi'_{i}\varphi'_{j}\chi' + 
c''_{1} \varphi''_{i}\varphi''_{j}\chi'') \bar{L}_{i} \phi_{1}e_{R\ j} + H.c. \nonumber  \\
& & + (c_{0}\varphi_{i}\varphi_{j}^{\ast} + c'_{0}\varphi'_{i}\varphi_{j}^{'\ast} 
+ c''_{0}\varphi''_{i}\varphi_{j}^{''\ast} + c\delta_{ij} )
 \bar{L}_{i} \phi_{2}\phi_{2}^{T}L_{j}^{c} + H.c. \nonumber \\
& & + D_{\mu}\varphi^{\ast} D^{\mu}\varphi + D_{\mu}\varphi^{'\ast}D^{\mu}\varphi'
  + D_{\mu}\varphi^{''\ast} D^{\mu}\varphi''  - V_{\varphi}   
\end{eqnarray}
which is assumed to be resulted from integrating out heavy particles. 
${\cal L}_{SM} $ denotes the lagrangian of the standard 
model. $\bar{L}_{i}(x) = (\bar{\nu}_{i}, \bar{e}_{i})_{L}$ 
(i=1,2,3) are the SU(2)$_{L}$ doublet leptons and $e_{R\ i}$ ($i=1,2,3$) are 
the three right-handed charged leptons. $A_{\mu}^{i}(x)t^{i}$ ($i=1,2,3$) 
are the $SO(3)_{F}$ gauge bosons with $t^{i}$ the $SO(3)_{F}$ generators and 
$g'_{3}$ is the corresponding gauge coupling constant. Here $\phi_{1}(x)$ and $\phi_{2}(x)$ 
are two Higgs doublets, $\varphi^{T}=(\varphi_{1}(x), \varphi_{2}(x), 
\varphi_{3}(x))$, $\varphi'^{T} = (\varphi'_{1}(x), \varphi'_{2}(x), 
\varphi'_{3}(x))$ and $\varphi''^{T} = (\varphi''_{1}(x), \varphi''_{2}(x), 
\varphi''_{3}(x))$ are three $SO(3)_{F}$ Higgs triplets, and $\chi(x)$, 
$\chi'(x)$ and $\chi''(x)$ are three singlet scalars. The couplings $c$, $c_{a}$, $c'_{a}$ and 
$c''_{a}$ ($a=0,1$) are dimensional constants. The structure of the above effective 
lagrangian can be obtained by imposing an additional U(1) symmetry \cite{YLW1}. 

   After the symmetry SO(3)$_{F}\times$SU(2)$_{L}\times$U(1)$_{Y}$ is broken 
down to the U(1)$_{em}$ symmetry, we obtain mass matrices of the neutrinos 
and charged leptons as follows
\begin{eqnarray}
(M_{e})_{ij} & = & m_{1} \frac{\hat{\sigma}_{i}\hat{\sigma}_{j}}{\sigma^{2}} 
+ m'_{1} \frac{\hat{\sigma}'_{i}\hat{\sigma}'_{j}}{\sigma^{'2}} + m''_{1} 
\frac{\hat{\sigma}''_{i}\hat{\sigma}''_{j}}{\sigma^{''2}} \nonumber \\
(M_{\nu})_{ij} & = &  m_{0} \frac{\hat{\sigma}_{i}
\hat{\sigma}_{j}^{\ast} + \hat{\sigma}_{j}\hat{\sigma}_{i}^{\ast} }{2\sigma^{2}}
+  m'_{0} \frac{\hat{\sigma}'_{i}\hat{\sigma}_{j}^{'\ast} + 
\hat{\sigma}'_{j}\hat{\sigma}_{i}^{'\ast}}{2\sigma^{'2}}  \nonumber \\  
& + & m''_{0} \frac{\hat{\sigma}''_{i}\hat{\sigma}_{j}^{''\ast} + 
\hat{\sigma}''_{j}\hat{\sigma}_{i}^{''\ast}}{2\sigma^{''2}} + m_{\nu} \delta_{ij}
\end{eqnarray}
where the mass matrices $M_{e}$ and $M_{\nu}$ are defined in the basis
 ${\cal L}_{M} = \bar{e}_{L}M_{e}e_{R} + 
\bar{\nu}_{L}M_{\nu}\nu^{c}_{L} + H.c. $. The constants $\hat{\sigma}_{i} = 
<\varphi_{i}(x)>$, $\hat{\sigma}'_{i} = <\varphi'_{i}(x)>$ and 
$\hat{\sigma}''_{i} = <\varphi''_{i}(x)>$
are the vacuum expectation values (VEVs) of the three Higgs triplets with 
$\sigma^{2} = \sum_{i=1}^{3}|\hat{\sigma}_{i}|^{2}$, $\sigma^{'2} = 
\sum_{i=1}^{3}|\hat{\sigma}'_{i}|^{2}$ and $\sigma^{''2} = 
\sum_{i=1}^{3}|\hat{\sigma}''_{i}|^{2}$. Here $m_{1}$, $m'_{1}$ and $m''_{1}$ as well as
$m_{\nu}$, $m_{0}$, $m'_{0}$ and $m''_{0}$ are mass parameters. 

  The Higgs potential for the $SO(3)_{F}$ Higgs triplets has the following general form before 
symmetry breaking 
\begin{eqnarray}
V_{\varphi} & = & \frac{1}{2} \mu^{2} (\varphi^{\dagger}\varphi ) + 
\frac{1}{2} \mu^{'2} (\varphi'^{\dagger}\varphi' ) + 
\frac{1}{2} \mu^{''2} (\varphi''^{\dagger}\varphi'' )  \nonumber \\
& + & \frac{1}{4}\lambda (\varphi^{\dagger}\varphi)^{2} + 
\frac{1}{4}\lambda' (\varphi'^{\dagger}\varphi')^{2} + 
\frac{1}{4}\lambda'' (\varphi''^{\dagger}\varphi'')^{2} \nonumber \\
& + & \frac{1}{2}\kappa_{1} (\varphi^{\dagger}\varphi )( \varphi'^{\dagger}\varphi') + 
\frac{1}{2}\kappa'_{1} (\varphi^{\dagger}\varphi )( \varphi''^{\dagger}\varphi'')  \\
& + & \frac{1}{2}\kappa''_{1} (\varphi'^{\dagger}\varphi' )( \varphi''^{\dagger}\varphi'') 
 + \frac{1}{2} \kappa_{2} (\varphi^{\dagger}\varphi')( \varphi'^{\dagger}\varphi ) \nonumber \\
& + &  \frac{1}{2} \kappa'_{2} (\varphi^{\dagger}\varphi'')( \varphi''^{\dagger}\varphi )
+ \frac{1}{2} \kappa''_{2} (\varphi'^{\dagger}\varphi'')( \varphi''^{\dagger}\varphi' )\  .  \nonumber 
\end{eqnarray}
For simplicity, we have omitted terms involving other Higgs fields since those terms 
will not change our conclusions and their effects may actually be absorbed into the 
redefinitions of the coupling constants. Also note that the above general form of the
Higgs potential is not changed after integrating of the heavy fermions though the other 
parts of Lagrangian become effective ones with dimension being larger than four.  
 
   As the $SO(3)_{F}$ flavor symmetry is treated to be a gauge symmetry, 
one can always express the complex $SO(3)_{F}$ Higgs triplet  
fields $\varphi_{i}(x)$ ($\varphi'_{i}(x)$, $\varphi''_{i}(x)$) in terms of  
three rotational fields $\eta_{i}(x)$ ($\eta'_{i}(x)$, $\eta''_{i}(x)$) and three 
amplitude fields $\rho_{i}(x)$ ($\rho'_{i}(x)$, $\rho''_{i}(x)$) ($i=1,2,3$), i.e., 
$\varphi(x) = O(x)\rho(x)$, $\varphi'(x) = O'(x)\rho'(x)$ and $\varphi''(x) = O''(x)\rho''(x)$
with  $O(x)\equiv e^{i\eta_{i}(x)t^{i}}$, $O'(x)\equiv e^{i\eta'_{i}(x)t^{i}}$, 
$O''(x)\equiv e^{i\eta''_{i}(x)t^{i}}$ $\in $ $SO(3)_{F}$ being the $SO(3)_{F}$ rotational fields. 
As discussed in refs.\cite{YLW1,YLW2,YLW0}, a complex Higgs triplet can be reparameterized 
in terms of the form  
\begin{equation} 
 \left( \begin{array}{c}
  \varphi_{1}(x) \\
  \varphi_{2}(x)   \\
  \varphi_{3}(x)   \\
\end{array} \right) = e^{i\eta_{i}(x)t^{i}} \frac{1}{\sqrt{2}}
\left( \begin{array}{c}
  \rho_{1}(x) \\
  i\rho_{2}(x)   \\
  \rho_{3}(x)   \\
\end{array} \right) 
\end{equation}
Similar forms are for $\varphi'(x)$ and  $\varphi''(x)$. $SO(3)_{F}$ gauge symmetry allows one to 
remove three degrees of freedom from nine rotational fields. 
Making $SO(3)_{F}$ gauge transformations: $(\varphi(x), \varphi'(x), \varphi''(x))
 \rightarrow O^{T}(x) (\varphi (x), \varphi'(x), \varphi''(x))$, and assuming that
only the amplitude fields get VEVs after spontaneous breaking of the $SO(3)_{F}$ 
flavor symmetry, namely  $<\rho_{i}(x)> = \sigma_{i}$, $<\rho'_{i}(x)> = \sigma'_{i}$ and  
$<\rho''_{i}(x)> = \sigma''_{i}$, we then obtain the following equations from minimizing the 
Higgs potential 
\begin{mathletters} 
\begin{eqnarray}
& & \omega^{2} \sigma_{i} 
+ \kappa_{2} \sum_{j=1}^{3}(\sigma_{j}\sigma'_{j}) \sigma'_{i} + 
\kappa'_{2} \sum_{j=1}^{3}(\sigma_{j}\sigma''_{j}) \sigma''_{i}  = 0  \\
& & \omega^{'2} \sigma'_{i}  
+ \kappa_{2} \sum_{j=1}^{3}(\sigma_{j}\sigma'_{j}) \sigma_{i} + 
\kappa''_{2} \sum_{j=1}^{3}(\sigma'_{j}\sigma''_{j}) \sigma''_{i}  = 0  \\
& & \omega^{''2} \sigma''_{i} + \kappa'_{2} \sum_{j=1}^{3}(\sigma_{j}\sigma'_{j}) \sigma_{i} + 
\kappa''_{2} \sum_{j=1}^{3}(\sigma'_{j}\sigma''_{j}) \sigma'_{i}  = 0 
\end{eqnarray} 
\end{mathletters}
with $\omega^{2} = \mu^{2} + \lambda \sigma^{2} + \kappa_{1} \sigma^{'2} + 
\kappa'_{1}\sigma^{''2} $, $\omega^{'2} = \mu^{'2} + \lambda' \sigma^{'2} + 
\kappa_{1} \sigma^{2} + \kappa''_{1}\sigma^{''2} $ and $\omega^{''2}= \mu^{''2} + 
\lambda'' \sigma^{''2} + \kappa'_{1} \sigma^{2} + \kappa''_{1}\sigma^{'2}$.
To find out possible constraints, it is useful to set $\sigma'_{i} = \xi_{i} \sigma_{i}$ 
and $\sigma'_{i} = \xi'_{i} \sigma_{i}$ for $\sigma_{i}\neq 0$ with $i=1,2,3$ as well as 
$\sigma^{'2} = \xi \sigma^{2}$ and $\sigma^{''2} = \xi' \sigma^{2}$.
When $\xi_{1} = \xi_{2} = \xi_{3}=\sqrt{\xi} $, the two $SO(3)_{F}$ 
Higgs triplets $\varphi(x)$ and $\varphi'(x)$ are parallel in the model and 
the introduction of the second Higgs triplet becomes trivial.
In the general and nontrivial cases, it is easy to check that when   
$\xi_{i}=\xi_{j} \neq \xi_{k}\equiv \xi_{i} - \xi_{0}$ and $\xi'_{i} \neq \xi'_{j}\equiv 
\xi'_{i} - \xi'_{0}$ with $i\neq j \neq k$, or 
$\xi_{i}\neq \xi_{k}$, $\xi_{j} \neq \xi_{k}$, $\xi'_{i} \neq \xi'_{k}$ and $\xi'_{j} \neq \xi'_{k}$
with $i\neq j \neq k$, one arrives at the strong constraints: 
$\sum_{i=1}^{3} \sigma_{i}\sigma''_{i} =0$, $\sum_{i=1}^{3} \sigma_{i}\sigma'_{i} =0$, 
$\omega^{2}= \mu^{2} + \lambda \sigma^{2} + \kappa_{1} \sigma^{'2} + \kappa'_{1}\sigma^{''2} = 0$ 
from the minimum conditions in eq.(5a), and 
$\sum_{i=1}^{3} \sigma'_{i}\sigma''_{i} =0$, $\sum_{i=1}^{3} \sigma_{i}\sigma'_{i} =0$ and
$\omega^{'2}=\mu^{'2} + \lambda' \sigma^{'2} + \kappa_{1} \sigma^{2} + \kappa''_{1}\sigma^{''2} =0$ 
from the minimum conditions in eq.(5b), as well as 
$\sum_{i=1}^{3} \sigma'_{i}\sigma''_{i} =0$, $\sum_{i=1}^{3} \sigma_{i}\sigma''_{i} =0$, 
and $\omega^{''2}=\mu^{''2} + \lambda'' \sigma^{''2} + \kappa'_{1} \sigma^{2} + 
\kappa''_{1}\sigma^{'2} = 0$ from the minimum conditions in eq.(5c). 
For convenience of discussions, we make, without lossing generality, the
convention that $\xi_{1}=\xi_{2} \neq \xi_{3}\equiv \xi_{1} - \xi_{0}$ and $\xi'_{1} \neq 
\xi'_{2} \equiv \xi'_{1} - \xi'_{0}$. 
Thus from the constraints: $\sum_{i=1}^{3}\sigma_{i}\sigma'_{i} = 
\sum_{i=1}^{3}\xi_{i}\sigma_{i}^{2}=0 $ 
and the expression $\sigma^{'2} = \xi \sigma^{2}$, i.e.,
$\xi_{1}(\sigma_{1}^{2} + \sigma_{2}^{2}) + \xi_{3} \sigma_{3}^{2} = 0$ and
$\xi_{1}^{2} (\sigma_{1}^{2} + \sigma_{2}^{2}) + \xi_{3}^{2} \sigma_{3}^{2} = \xi \sigma^{2} $,
we obtain the solutions  
\begin{equation}
\xi = \xi_{1}(\xi_{0}-\xi_{1}), \  \  \xi - \xi_{1}^{2}\tan^{2}\theta_{2}  =0
\end{equation}
with $\tan^{2}\theta_{2}=\sigma_{12}^{2}/\sigma_{3}^{2}$ and 
$\sigma_{12}^{2}= \sigma_{1}^{2} + \sigma_{2}^{2}$.  From the constraints:  
$\sum_{i=1}^{3}\sigma_{i}\sigma''_{i} = \sum_{i=1}^{3}\xi'_{i}\sigma_{i}^{2} = 0 $,  
$\sum_{i=1}^{3}\sigma'_{i}\sigma''_{i} = \sum_{i=1}^{3}\xi_{i}\xi'_{i}\sigma_{i}^{2} = 0 $ 
and the experession $\sigma^{''2} = \xi' \sigma^{2}$, i.e., 
$\xi_{1}(\xi'_{1}\sigma_{1}^{2} + \xi'_{2}\sigma_{2}^{2}) + \xi_{3}\xi'_{3} \sigma_{3}^{2} = 0$,
$\xi'_{1}\sigma_{1}^{2} + \xi'_{2}\sigma_{2}^{2} + \xi'_{3} \sigma_{3}^{2} = 0$ and
$\xi_{1}^{'2} \sigma_{1}^{2} + \xi_{2}^{'2} \sigma_{2}^{2} + \xi_{3}^{'2} 
\sigma_{3}^{2} = \xi' \sigma^{2} $, we yield the following solutions by combining 
the solutions in eq.(6),  
\begin{eqnarray}
& & \xi' = \xi'_{1} (\xi'_{0} -\xi'_{1})/(1 + \xi_{1}^{2}/\xi) , \nonumber \\
& & \xi'(1 + \xi_{1}^{2}/\xi) - \xi_{1}^{'2}\tan^{2}\theta_{1}  = 0, \  \  \xi'_{3}=0
\end{eqnarray}
with $\tan^{2}\theta_{1} = \sigma_{1}^{2}/\sigma_{2}^{2}$. 

   In addition, one needs to further check the minimum conditions directly from 
the Higgs potential at the minimizing point. It is not difficult to find that 
\begin{eqnarray}
V_{\varphi}|_{min} & = & -\sigma^{4}(\lambda + \lambda'\xi^{2} + \lambda''\xi^{'2} \nonumber \\ 
& + & 2\kappa_{1}\xi + 2\kappa'_{1}\xi' + 2\kappa''_{1}\xi\xi')/4
\end{eqnarray}
which shows that to have a global minimum potential energy $V_{\varphi}|_{min}$ for 
varying $\xi$ and $\xi'$, the values of $\xi$ and $\xi'$ are required to be maximal 
for positive coupling constants $\lambda$'s and $\kappa$'s. From such a requirement, it is seen 
that for the given $\xi_{0}$ and $\xi'_{0}$ in eqs.(6) and (7), the maximum conditions for 
$\xi$ and $\xi'$ lead to the solutions $\xi_{1}= \xi_{0}/2 = \sqrt{\xi} = \xi_{2} = -\xi_{3}$ 
and $\xi'_{1} = \xi'_{0}/2 = \sqrt{2\xi'} = -\xi'_{2}$. Summarizing all of the above results, 
we arrive at the relations
\begin{eqnarray}
& & \sigma'_{1}= \sqrt{\xi} \sigma_{1}, \    \sigma'_{2}= \sqrt{\xi} \sigma_{2}, 
\   \sigma'_{3}= -\sqrt{\xi} \sigma_{3}, \nonumber \\ 
& & \sigma''_{1}= \sqrt{2\xi'} \sigma_{1}, \  \  \sigma''_{2}= -\sqrt{2\xi'} \sigma_{2}, 
\  \  \sigma''_{3}= 0,  \\ 
& &  \sigma_{3}^{2} = \sigma_{1}^{2} + \sigma_{2}^{2} =2\sigma_{1}^{2}= \sigma^{2}/2, 
\  i.e.,  \  \theta_{1} = \theta_{2} = \pi/4  \nonumber 
\end{eqnarray}
Noticing that from the constraints: $\omega^{2} = 0 $, $\omega^{'2} = 0 $ and $\omega^{''2} = 0 $, 
one can yield $\sigma^{2}$, $\sigma^{'2}$ and $\sigma^{''2}$, namely $\xi$ and $\xi'$, as 
functions of the coupling constants in the Higgs potential. Thus the VEVs are completely determined 
by the Higgs potential.

  It is remarkable that with these relations the mass matrices of the neutrinos and 
charged leptons are greatly simplified to the following nice forms
\begin{equation}
M_{e}  =  \frac{m_{1}}{2}\left( \begin{array}{ccc}
  \frac{1}{2} & \frac{1}{2}i & \frac{1}{\sqrt{2}}  \\
   \frac{1}{2}i & -\frac{1}{2} &  \frac{1}{\sqrt{2}}i \\
  \frac{1}{\sqrt{2}} & \frac{1}{\sqrt{2}}i & 1 \\ 
\end{array} \right) 
 +  \frac{m'_{1}}{2}\left( \begin{array}{ccc}
 \frac{1}{2} & \frac{1}{2}i & -\frac{1}{\sqrt{2}}  \\
   \frac{1}{2}i & -\frac{1}{2} &  -\frac{1}{\sqrt{2}}i \\
  -\frac{1}{\sqrt{2}} & -\frac{1}{\sqrt{2}}i & 1 \\ 
\end{array} \right) + \frac{m''_{1}}{2} \left( \begin{array}{ccc}
  1 & i & 0  \\
  i & -1  & 0 \\
 0 & 0 &  0 \\ 
\end{array} \right)
\end{equation}
and
\begin{equation}
M_{\nu} = \hat{m}_{\nu}\left( \begin{array}{ccc}
  1  & 0 & \frac{1}{\sqrt{2}}\hat{\delta}_{-}  \\
  0 & 1   &  0 \\
  \frac{1}{\sqrt{2}}\hat{\delta}_{-} & 0
 & 1 + \hat{\Delta}_{-}  \\ 
\end{array} \right)
\end{equation}
with $\hat{m}_{\nu} = m_{\nu}(1+ \Delta_{+})$,  $\hat{\delta}_{-} = \delta_{-}/(1+ \Delta_{+})$ 
and  $\hat{\Delta}_{-} = \Delta_{-}/(1+ \Delta_{+})$. 
Here $\Delta_{\pm} = (\delta_{+} \pm \delta_{0})/2$, $\delta_{0} = m''_{0}/m_{\nu}$ 
and $\delta_{\pm} = (m_{0}\pm m'_{0})/2m_{\nu}$ . 

  It is more remarkable that the mass matrix $M_{e}$ can be diagonalized by a unitary 
bi-maximal mixing matrix $U_{e}$ via $D_{e} = U_{e}^{\dagger} M_{e} U_{e}^{\ast}$ with
\begin{equation}
U_{e}^{\dagger}=\left( \begin{array}{ccc}
  \frac{1}{\sqrt{2}}i & -\frac{1}{\sqrt{2}} & 0  \\
 \frac{1}{2} & -\frac{1}{2}i & -\frac{1}{\sqrt{2}} \\
 \frac{1}{2} & -\frac{1}{2}i & \frac{1}{\sqrt{2}}  \\
\end{array} \right)
\end{equation}
and $D_{e} = diag. (m_{e}, m_{\mu}, m_{\tau})$. 
Here $m_{e} = -m''_{1}$, $m_{\mu} =  m'_{1}$ and  $m_{\tau} =  m''_{1}$ are
the charged lepton masses. The neutrino mass matrix can be easily diagonalized 
by an orthogonal matrix $O_{\nu}$ via $O_{\nu}^{T}M_{\nu}O_{\nu}$ with 
$(O_{\nu})_{13}=\sin \theta_{\nu}\equiv  s_{\nu}$ and 
$\tan2\theta_{\nu} = \sqrt{2}\hat{\delta}_{-}/\hat{\Delta}_{-}$.
Thus the CKM-type lepton mixing matrix $U_{LEP}$ that   
appears in the interaction term 
${\cal L}_{W} = \bar{e}_{L}\gamma^{\mu}U_{LEP} \nu_{L} W_{\mu}^{-} + H.c. $
is given by $U_{LEP} = U_{e}^{\dagger}O_{\nu}$. Explicitly, one has 
\begin{equation}
U_{LEP}  = \left( \begin{array}{ccc}
  \frac{1}{\sqrt{2}}ic_{\nu} & -\frac{1}{\sqrt{2}} & \frac{1}{\sqrt{2}}is_{\nu}  \\
 \frac{1}{2}c_{\nu}+ \frac{1}{\sqrt{2}}s_{\nu} & -\frac{1}{2}i & 
\frac{1}{2}s_{\nu}-\frac{1}{\sqrt{2}}c_{\nu} \\
 \frac{1}{2}c_{\nu}-\frac{1}{\sqrt{2}}s_{\nu} & -\frac{1}{2}i & 
\frac{1}{2}s_{\nu}+ \frac{1}{\sqrt{2}}c_{\nu}  \\
\end{array} \right).
\end{equation}
The three neutrino masses are found to be
\begin{eqnarray}
m_{\nu_{e}} & = &  \hat{m}_{\nu}[1 -  
(\sqrt{\hat{\Delta}_{-}^{2} + 2 \hat{\delta}_{-}^{2}} - \hat{\Delta}_{-} )/2\  ]
\nonumber \\
m_{\nu_{\mu}} & = &  \hat{m}_{\nu}   \\
m_{\nu_{\tau}} & = & \hat{m}_{\nu}[1 + \hat{\Delta}_{-} + 
(\sqrt{\hat{\Delta}_{-}^{2} + 2 \hat{\delta}_{-}^{2}} - \hat{\Delta}_{-} )/2\  ] . \nonumber 
\end{eqnarray}

  The similarity between the Higgs triplets $\varphi(x)$ and $\varphi'(x)$
naturally motivates us to consider an approximate (and softly broken) permutation 
symmetry between them. This implies that $m_{0}\simeq m'_{0}$. As a consequence, 
one has $|\hat{\delta}_{-}|<< 1$. To a good approximation, the the mass-squared 
differences are given by  $\Delta m_{\mu e}^{2} \equiv m_{\nu_{\mu}}^{2} - 
m_{\nu_{e}}^{2} \simeq  \hat{m}_{\nu}^{2}\hat{\Delta}_{-}(\hat{\delta}_{-}/\hat{\Delta}_{-})^{2}$ 
and  $\Delta m_{\tau\mu}^{2} \equiv  m_{\nu_{\tau}}^{2} - m_{\nu_{\mu}}^{2} \simeq 
 \hat{m}_{\nu}^{2} \hat{\Delta}_{-}(2 + \hat{\Delta}_{-})$,  which leads to the following 
approximate relation
\begin{equation}
 \frac{\Delta m_{\mu e}^{2}}{\Delta m_{\tau\mu}^{2}}
\simeq \left(\frac{\hat{\delta}_{-}}{\sqrt{2}\hat{\Delta}_{-}}\right)^{2} 
\simeq s^{2}_{\nu}= 2|U_{e3}|^{2}  << 1 \  .
\end{equation}
The allowed range of the ratio is $\Delta m_{\mu e}^{2}/\Delta m_{\tau\mu}^{2}
 \sim 10^{-2}-10^{-8}$. Here the large value is for matter-enhanced MSW solution\cite{MSW,MSW1} 
with large mixing angle\cite{LAMSW,LAMSW1} and the small value for the vacuum 
oscillation solutions\cite{VO,VO1}\footnote{See also  V. Berezinsky, 
G. Fiorentini and M. Lisia, hep-ph/9811352.}. With the hierarchical feature in $\Delta m^{2}$,  
formulae for the oscillation probabilities in vacuum are greatly simplified to be
\begin{eqnarray}
& & P_{\nu_{e}\rightarrow \nu_{e}}|_{solar} \simeq  1 - 
\sin^{2}(\frac{\Delta m_{\mu e}^{2}L}{4E}) \nonumber \\
& & P_{\nu_{\mu}\rightarrow \nu_{\mu}}|_{atm.} \simeq 1 -
\sin^{2}(\frac{\Delta m_{\tau\mu}^{2}L}{4E}) \\
& & P_{\nu_{\beta}\rightarrow \nu_{\alpha}} \simeq 4|U_{\beta 3}|^{2}|U_{\alpha 3}|^{2}
\sin^{2}(\frac{\Delta m_{\tau\mu}^{2}L}{4E}) \nonumber \\
& & P_{\nu_{\mu}\rightarrow \nu_{e}}/P_{\nu_{\mu}\rightarrow \nu_{\tau}}|_{atm.}
\simeq (\Delta m_{\mu e}^{2}/\Delta m_{\tau\mu}^{2}) << 1 \   . \nonumber 
\end{eqnarray}
This may present the simplest scheme for reconciling both 
solar and atmospheric neutrino fluxes via oscillations of three neutrinos. 

  When going back to the weak gauge and charged-lepton mass basis, 
the neutrino mass matrix gets the following interesting form 
\begin{equation}
M_{\nu}/\hat{m}_{\nu} \simeq \left( \begin{array}{ccc}
  0 & \frac{1}{\sqrt{2}}i & \frac{1}{\sqrt{2}}i  \\
   \frac{1}{\sqrt{2}}i & \frac{1}{2} &  - \frac{1}{2}  \\
  \frac{1}{\sqrt{2}}i &  - \frac{1}{2} &  \frac{1}{2}  \\ 
\end{array} \right) + \frac{\hat{\delta}_{-}}{2}\left( \begin{array}{ccc}
 0 & -\frac{1}{\sqrt{2}}i & \frac{1}{\sqrt{2}}i  \\
   -\frac{1}{\sqrt{2}}i & -1 &  0  \\
  \frac{1}{\sqrt{2}}i & 0 &  1 \\  
\end{array} \right) + \frac{\hat{\Delta}_{-}}{2}\left( \begin{array}{ccc}
 0 & 0 & 0  \\
   0 & 1 &  -1  \\
  0 & -1 &  1 \\  
\end{array} \right).
\end{equation}
As $(M_{\nu})_{ee} = 0$, the neutrinoless double beta decay\cite{DB} 
is forbiden in the model. Thus the neutrino masses can be approximately 
degenerate and large enough  ( $\hat{m}_{\nu}= O(1)$ eV) to play a 
significant cosmological role. 

 The mass matrix of gauge fields $A_{\mu}^{i}$ is found to be
\begin{equation}
 M_{F}^{2} = \frac{m_{F}^{2}}{3}\left( \begin{array}{ccc}
  2(\xi_{+} + \xi')  & 0  & -\sqrt{2}\xi_{-} \\ 
    0 & 3\xi_{+} + \xi' & 0  \\ 
 -\sqrt{2}\xi_{-} & 0  &  3\xi_{+} + \xi'   \\ 
\end{array} \right)  
\end{equation} 
with $ m_{F}^{2}= 3g^{'2}_{3}\sigma^{2}/8$ and $\xi_{\pm}=(1 \pm \xi)/2$. 
This mass matrix is diagonalized 
by an orthogonal matrix $O_{F}$ via $O_{F}^{T}M_{F}^{2}O_{F}$.  
Denoting the physical gauge fields as $F_{\mu}^{i}$, 
we then have $A_{\mu}^{i} = O_{F}^{ij}F_{\mu}^{j}$, i.e., 
\begin{equation} 
\left( \begin{array}{c} A_{\mu}^{1} \\ A_{\mu}^{2} \\ A_{\mu}^{3} \\ \end{array} 
\right)  = \left( \begin{array}{ccc} c_{F} & 0  & -s_{F} \\ 0 & 1 & 0  \\ s_{F}  
&  0 & c_{F}  \\ \end{array} \right) \left( \begin{array}{c} F_{\mu}^{1} \\ 
F_{\mu}^{2} \\ F_{\mu}^{3} \\ \end{array} \right) 
\end{equation} 
with $s_{F}\equiv \sin \theta_{F}$ and 
$\tan 2\theta_{F}  = 2\sqrt{2}\xi_{-}/(\xi_{+} - \xi')$.  
Masses of the three physical gauge bosons $F_{\mu}^{i}$ are given by
\begin{eqnarray}
m_{F_{1}}^{2} & = &   m_{F}^{2}( 5\xi_{+} + 3\xi'  - 
 \sqrt{(\xi_{+} -\xi')^{2} + 8\xi_{-}^{2}}\  )/6 \nonumber \\
m_{F_{2}}^{2} & = &   m_{F}^{2}( \xi_{+} + \xi'/3) \\
m_{F_{3}}^{2} & = &   m_{F}^{2} ( 5\xi_{+} + 3\xi' + 
 \sqrt{(\xi_{+} -\xi') + 8\xi_{-}^{2}} \  ) /6  \nonumber 
\end{eqnarray}
In the physical mass basis, we have for gauge interactions 
\[
{\cal L}_{F} =  \frac{g'_{3}}{2} F_{\mu}^{i}\left(\bar{\nu}_{L}t^{j}
O_{F}^{ji}\gamma^{\mu}\nu_{L}
+ \bar{e}_{L} V_{e}^{i}\gamma^{\mu}e_{L} - \bar{e}_{R} V_{e}^{i \ast}
\gamma^{\mu}e_{R}\right) 
\]
with $V_{e}^{i} = U_{e}^{\dagger}t^{j}U_{e}O_{F}^{ji}$. Explicitly, we find 
\begin{eqnarray} 
V_{e}^{1} & = & \left( \begin{array}{ccc}
  c_{F} & i\frac{1}{2}s_{F} & -i\frac{1}{2}s_{F} \\
   -i\frac{1}{2}s_{F} & \frac{1}{2}c_{F} + \frac{1}{\sqrt{2}}s_{F} &  \frac{1}{2}c_{F}  \\
  i\frac{1}{2}s_{F} &  \frac{1}{2}c_{F} & \frac{1}{2}c_{F} -\frac{1}{\sqrt{2}}s_{F}  \\ 
\end{array} \right) \nonumber \\
 V_{e}^{2} & = & \left( \begin{array}{ccc}
  0 & \frac{1}{2} & -\frac{1}{2} \\
   \frac{1}{2} & 0 &  i\frac{1}{\sqrt{2}}  \\
 -\frac{1}{2} &  -i\frac{1}{\sqrt{2}} & 0 \\ 
\end{array} \right)  \\
V_{e}^{3} & = & \left( \begin{array}{ccc}
  -s_{F} & i\frac{1}{2}c_{F} & -i\frac{1}{2}c_{F} \\
   -i\frac{1}{2}c_{F} & -\frac{1}{2}s_{F} + \frac{1}{\sqrt{2}}c_{F} &  -\frac{1}{2}s_{F}  \\
  i\frac{1}{2}c_{F} &  -\frac{1}{2}s_{F} & -\frac{1}{2}s_{F} -\frac{1}{\sqrt{2}}c_{F}  \\ 
\end{array} \right).  \nonumber
\end{eqnarray}
Thus the $SO(3)_{F}$ gauge interactions allow lepton flavor violating process 
$\mu \rightarrow 3e$, its branch ratio is found
\begin{equation}
Br(\mu \rightarrow 3e) = \left(\frac{v}{\sigma}\right)^{4}
 \frac{2\xi_{-}^{2}}{[(3\xi_{+} + \xi')(\xi_{+} + \xi') - \xi_{-}^{2} ]^{2} } 
\end{equation}
with $v=246$GeV. For $\sigma \sim 10^{3} v$, the branch ratio could be very close to 
the present experimental upper bound $Br(\mu \rightarrow 3e) < 1 \times 10^{-12}$ \footnote{
See summary talk by D. Bryman at ICHEP98, Vancouver, Canada, 1998.}.
Thus when taking the mixing angle $\theta_{F}$ and the coupling constant $g'_{3}$ for the 
$SO(3)_{F}$ gauge bosons to be at the same order of magnitude as those for the 
electroweak gauge bosons, we find that masses of the SO(3)$_{F}$ gauge bosons 
are at the order of magnitudes $m_{F_{i}} \sim 10^{3} m_{W}\simeq 80 $ TeV. 

  In summary, we have constructed the SO(3)$_{F}$ gauge model with three Higgs triplets. 
It has been shown that after spontaneous breaking of SO(3)$_{F}$ flavor symmetry, the intriguing 
bi-maximal neutrino mixing scenario can be naturally obtained as a nontrival solution 
of the equations which arise from minimizing the Higgs potential $V_{\varphi}$.  
The hierarchy between the neutrino mass-squared differences can be naturally resulted from 
an approximate permutation symmetry between two Higgs triplets $\varphi$ and $\varphi'$. 
Nearly degenerate neutrino masses also come out naturally and allow to be large enough 
to play an essential role in the evolution of the large-scale structure of the universe. 
Though the neutrinoless double beta decay is forbiden, the SO(3)$_{F}$ gauge interactions 
may lead to interesting phenomena on lepton-flavor violations. Finally, we would like to point 
out that such a scheme remains stable\cite{YLW3} after renormalization group effects 
as the SO(3)$_F$ symmetry breaking scale in our scheme is much lower than the grand unification scale.

 {\bf Ackowledgments}: This work was supported in part by the NSF of China under the 
grant No. 19625514.

%\begin{thebibliography}{99}

\end{document}